# A Millimeter Wave MIMO Testbed for 5G Communications

Tian Hong Loh, David Cheadle, Philip Miller

National Physical Laboratory, Teddington, United Kingdom

*Abstract* — This paper presents a 2 × 2 millimeter wave (mm-wave) multiple-input-multiple-output (MIMO) testbed that operates at around 30 GHz. The link assessment of the system operating at 26.25 GHz was carried out on a test bench, with a short communication distance between the transmitting and receiving antennas. A user-programmable, reconfigurable and real-time signal processing field-programmable gate arrays (FPGAs)-based software defined radio (SDR) system was employed as part of the testbed to validate the system-level performance for a downlink time division long-term evolution (TD-LTE) duplex scheme. Constellation diagram for quadrature phase shift keying (QPSK) digital modulation were acquired while the testbed was operating at 30 GHz. The testbed could be employed for the development of signal test, communication algorithm and measurement metrology for 5G communications.

## I. INTRODUCTION

Although fourth-generation (4G) systems have been generally fulfilling current demands, it is beyond their capacity to satisfy all the likely future demands and system requirements. According to the executive summary of Networld2020 [1], the demands on fifth-generation (5G) systems, are likely to be up to a 1000 times greater than current 4G systems. The major demands will be a larger available throughput, a 1 ms service-level latency, 90% less energy consumption, and 10 times longer battery lifetime, all combined with a seamless coverage experience. 5G systems will need to evolve beyond the current state of the art in order to accommodate these demands. The 5G communication industry will face a wide variety of issues to break through the current limitations.

Multiple-input-multiple-output (MIMO) antennas have a significant role in 4G and 5G communications, both to increase system spectral efficiency and to increase energy efficiency [2]-[7]. 5G cellular communications are expected to be standardized by 2020 [8] when the candidate frequency bands for 5G technology will be defined. These are expected to include, millimeter waves (mm-waves), which will bring an evolution in the capabilities of cellular communications [6]-[12]. However, there is still a debate about what mm-wave frequencies will be made available with the entire spectrum up to 100 GHz being considered [6]-[8].

In this study a 2 × 2 mm-wave MIMO testbed was built using a four field-programmable gate arrays (FPGA)-based vector signal transceiver (VST) system modules with a real-time signal processing software defined radio (SDR) capability [5]. This was designed to operate at around 30 GHz. It was built to gain experience which could be fed into the design of future mm-wave massive MIMO testbeds. This paper is organized as follow: Section II describes the measurement setup, Section III presents some measurement results, and finally, conclusions are drawn in Section IV.

## II. SYSTEM DESIGN

As depicted in Fig. 1, the mm-wave MIMO testbed was setup on a test bench within the control room of an anechoic chamber with metallic wall. It was located on top of low reflectivity Rohacell and polystyrene foams with radio absorbing materials beneath them. It consisted of a signal generator, a frequency distribution unit, mixers, doublers, filters, power dividers, amplifiers, VSTs and suitable interconnects. Figure 2 shows further schematic details of the testbed.

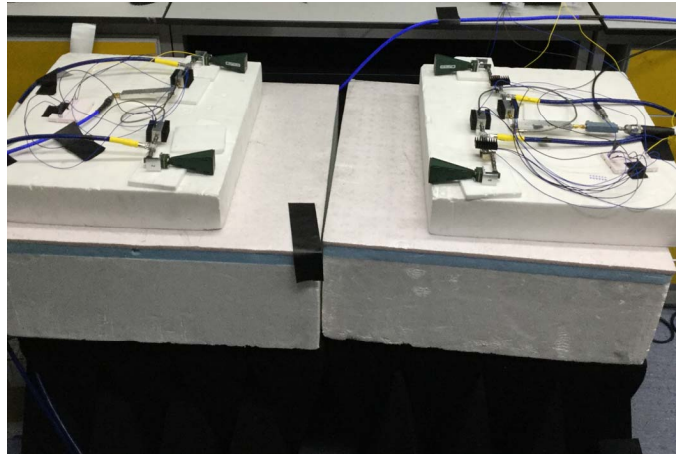

Fig. 1. Measurement setup of the mm-wave MIMO testbed on a test bench.

Four NI PXIe 5644R VST modules were employed.in the test set-up. These modules combine a Vector Signal Generator (VSG) and a Vector Signal Analyzer (VSA). They also possess Field Programmable Gate Arrays (FPGA), which are user programmable using LabVIEW to allow for real-time signal processing and control. The modules have has independent local oscillators (LOs) for both RF input and RF output with a frequency coverage from 65 MHz to 6 GHz.

As depicted in Fig. 2(a), a DC feedback circuit is used to stabilize the LO amplitude of the frequency distribution unit. Using a frequency doubler the LO signal, typically set at 12.5 GHz, is multiplied up to 25 GHz. This is mixed with a 5 GHz RF signal from the VST to achieve an output signal of

30 GHz. A filter is used to suppress the image frequency of 20 GHz.

For a two channel system, the transmit end uses a power splitter to allow the LO to drive both mixers. The same techniques was employed to share the LO drive between the mixers at the receive end (see Fig. 2(b)). The radiating elements and the receiving elements consist of two pairs of standard gain horns (Fig. 1).

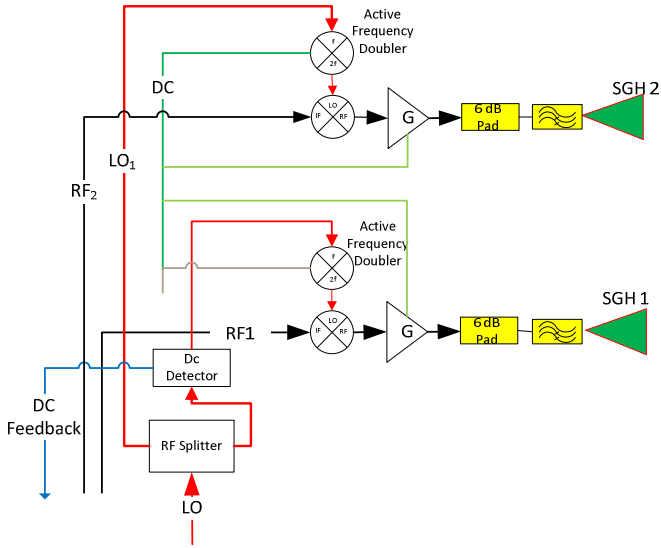

(a)

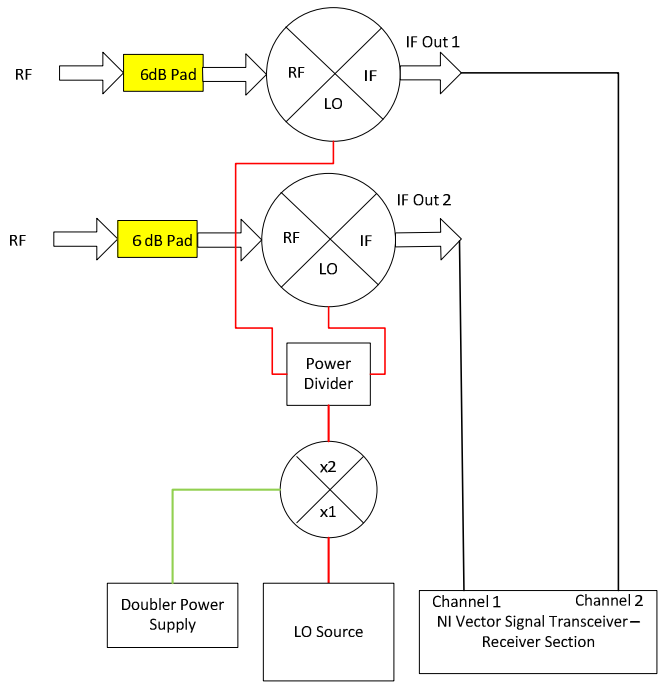

(b)

Fig. 2. System Layout: (a) Transmit system; (b) Receive system.

## III. LINK ASSESSMENT VALIDATION

As depicted in Fig. 1, the mm-wave MIMO testbed was setup on a test bench and the distance between the transmitting and receiving antennas was 38.3 cm. Figure 3 shows the measured link assessment comparison of the IF and RF signals between the transmitting and receiving ends while the LO frequency is 25 GHz. A calibrated Agilent E4440A performance spectrum analyzer (PSA) was used. The results shows that the loss on the IF and RF links between the transmitting and receiving ends is about 17 dB and 16.5 dB, respectively. Note that the difference in noise level in shown Fig. 3(a) is due to the different chosen resolution bandwidths.

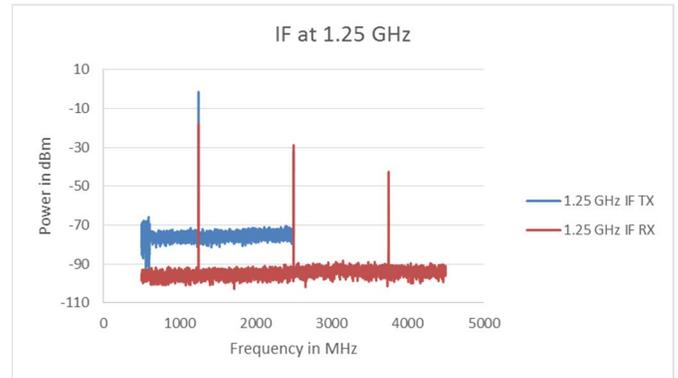

(a)

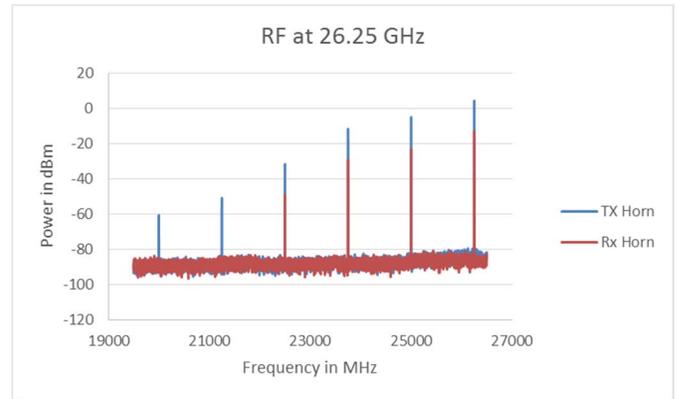

(b)

Fig. 3. Comparison of the measured power levels for the IF and RF signals between the transmitting and receiving ends: (a) IF at 1.25 GHz ('RF1' signal in Fig. 2(a) and 'IF Out 1' signal in Fig. 2(b)); (b) RF at 26.25 GHz (signal before SGH1 horn at the transmitting end and signal before the pad at the receiving end).

## IV. SYSTEM-LEVEL PERFORMANCE VALIDATION

For the system-level validation, a 2 × 2 TD-LTE MIMO system was employed using Quadrature Phase Shift Keying (QPSK) digital modulation with a transmitting frequency of 30 GHz. Figure 4 shows the measured system constellation plots of one complete LTE data frame with and without a metal plate. The metal plate was located very close to the transmitting end to block the direct link between the transmitting and receiving ends. The results indicate that, without the metal plate blocking (i.e. line-of-sight (LOS) scenario), the MIMO system shows a good communication link between the two ends whereas the expected deviation from the perfect QPSK happens when the metal plate is used (i.e. non-line-of-sight (NLOS) scenario). Note that multipath reflections exist for both LOS and NLOS scenarios within the metallic wall control room. In the LOS the direct path dominate whereas in the NLOS scenario the multipath dominates and there is potential deflection of surface current induced on the metallic plate.

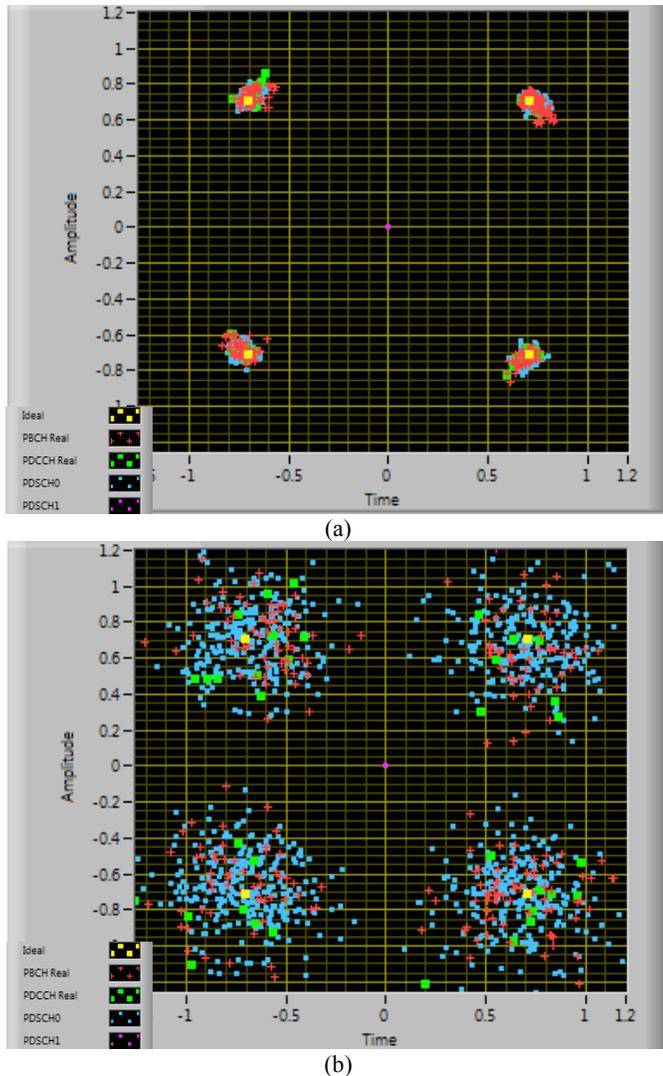

Fig. 4. The constellation diagram of the 2 x 2 MIMO transceiver system: (a) no metal plate blocking; (b) with metal plate blocking.

## V. CONCLUSIONS

This paper has presented a 2 × 2 mm-Wave MIMO testbed system operating at 30 GHz, which could be used for 5G communications studies. The system offers a degree of flexibility that enables the exploitation, especially the configurability to adapt the emergence of new standards and protocols. An example downlink TD-LTE duplex scheme was employed. The preliminary results of system verification on link assessment at 26.25 GHz and system-level link performance at 30 GHz are shown. The testbed could be employed for the development of signal test, communication algorithm and measurement metrology for 5G communications.

ACKNOWLEDGEMENT

The results in this paper come from the project MET5G – Metrology for 5G communications. This project has received funding from the EMPIR programme co-financed by the Participating States and from the European Union's Horizon 2020 research and innovation programme. The authors would like to thank Georgios Tsalavoutis of National Instruments for his help on VST hardware synchronization, and thank Yunsong Gui, Haowen Wang and Dr Fei Qin of Chinese Academy of Sciences for providing help over the development over the LabVIEW software codes.